\documentclass[12pt]{article}
\usepackage{graphicx}
\usepackage{amssymb,amsmath,epsfig}
\usepackage{epstopdf}
\usepackage{epsf}
\begin{document}

\begin{center}
\Large\textbf {Influence of pressure on magnetic phase transitions in the ferromagnetic superconductor UGe$_{2}$ - phenomenological approach}

\medskip

\normalsize
\textbf{Diana V. Shopova}
\end{center}
\medskip

\begin{center}
\normalsize
Institute of Solid State Physics, Bulgarian Academy of Sciences, \\
1784 Sofia, Bulgaria, sho@issp.bas.bg
\end{center}

 \medskip

\begin{abstract}
We propose a thermodynamic model of free energy expansion up to eighth order of magnetisation to describe the complex magnetic phase transitions in ferromagnetic superconductor UGe$2$. The model successfully describes transitions between ordered phases which take place without changing of magnetic structure but only of magnetisation which is the case of UGe$_2$ where  two magnetic phases with same structure but with different magnitude  occur with decreasing of temperature. We consider the influence of pressure on magnetic transitions in the simplest form of including it only in pressure dependence of Curie temperature for the transition between the paramagnetic and the magnetic phase with the lower magnetic moment FM1. Our results show the for pressures lower than some limiting pressure the transition between low-and high-magnetisation phase remains of crossover type. Above this limiting pressure this transition changes to real thermodynamic transition of first order.
\end{abstract}

\textbf{Keywords}: magnetic phase transition, ferromagnetic superconductor, Landau free energy\\

\textbf{PACS}: 74.20.De, 74.20.Rp, 74.40-n, 74.78-w

\section{\label{1} Introduction}

The uranium-based ferromagnetic superconductor UGe$_2$, ~\cite{Pfleiderer:2009},~\cite{Saxena:2000} is the first discovered one and there the superconductivity  occurs only under pressure in the pressure interval between $\sim$ 1GPa and critical pressure $P_c$ = 1.5 GPa where both superconductivity and magnetic order disappear. The magnetic phases are two with the same crystal nd magnetic structure but different magnitude of magnetisation. The high temperature  magnetic phase denoted by FM1 occurs by second order phase transition and Curie temperature at $P=0$  is 52.2K. With decreasing of temperature second ferromagnetic phase appears with larger magnetic moment at temperature of 30 K and at P =0.  The magnetic moment $M$  of FM1 phase experimentally found is approximately 0.9 $\mu_B$/U, while the magnetic moment of FM2 phase is 1.41 $\mu_B$/U ~\cite{Huxley:2015}. Recent summarised phase diagram with experimental data can be found in \cite{Tateiwa:2018}. The  second order phase transition to FM1 with the growing of pressure  turns into first order at tricritical point T$_{CP} \simeq 22 $ K,and  $P_{CP} \approx 1.42$ GPa .\\
It is experimentally found that at ambient  pressure the transition between FM1 and FM2 ferromagnetic phases is of crossover type and only with increasing pressure there is a true first order phase transition close to the appearance of superconductivity with a maximum transition temperature of $\approx 0.7$ K at P$_x=$ 1.2 GPa. In the pressure interval of superconductivity existence,  the magnetic  moment is $1 \mu_B/U$ .
It is established that UGe$_2$ has very strong uniaxial magnetic anisotropy~\cite{Onuki:1992}, and at some earlier stages of study, it has been considered that 3D Ising model can well describe its magnetic properties - see, for example,~\cite{Shopova:2009}, and for recent review ~\cite{Aoki:2019}. It is experimentally found and theoretically  justified~\cite{Tateiwa:2014} that the  critical exponents for UGe$_2$ do not belong to any known universality class including 3D Ising model and anisotropic next nearest neighbor 3D Ising model. One of the reasons for this lays in the dualism of 5f-electrons in UGe$_2$ where two subsets of itinerant and localised ones exist. This fact is experimentally established by positive muon relaxation measurements where the  itinerant magnetic moment at ambient pressure is found to be $\approx $ 0.02$ \mu_B$ ~\cite{Yaouanc:2002}. Later this dualism of 5f-electrons is confirmed by magnetic, transport and specific heat experiments, see, for example, ~\cite{Troc:2012}. \\ Understanding of magnetic phase transitions from microscopic point of view is important for understanding the appearance of superconductivity - for review, see for example ~\cite{Huxley:2015}, but there is now up to now consensus on this problem. Recently, an estimate is made on the basis of Takahashi's spin fluctuation theory about the itinerancy of 5f-electron systems in actinides~\cite{Tateiwa:2017}. The comparison of this approach with the experimental data shows that UGe$_2$ within the criterion used in the paper is an intermediate case between strongly localised and itinerant 5f-electron system. In the paper~\cite{Tateiwa:2018}, measurements of magnetization are performed at high pressure for UGe$_2$ with the aim to make clear the correlation between the superconductivity and pressure-enhanced ferromagnetic fluctuations. From the experimental data a conclusion is made that at lower pressure the ferromagnetic state is more itinerant in FM1 compared to FM2. \\
Usually microscopic models are focused on the phase transition between FM1 and FM2 around the critical end point, where superconductivity occurs under pressure ~\cite{Fidrysiak:2019}. At  pressures $0\leq P< P_x$  the nature of FM1 to FM2 transition is considered a crossover, and  in this pressure interval  there is small number of systematic measurements. Different theoretical explanations of this crossover transition exist.   In earlier studies performed with Stoner theory of itinerant electrons, it is proposed that the electronic density of states has two very closely placed peaks near the low-dimensional anisotropic Fermi surface, whose topology  changes by the appearance of magnetisation ~\cite{Sandeman:2002}.\\
We have proposed in previous papers \cite{Shopova:2020; Shopova:2021} that phenomenologically the magnetic transitions in UGe$_2$ fall in the class of isostructural transition, for which the transitions between ordered phases occur without the change of crystal or  magnetic structure, and only some physical quantity is changed as the magnitude of magnetisation in UGe$_2$. It was proposed by~\cite{Gufan:1978} and later on generalised see, for example \cite{Izyumov:1984}, \cite{Toledano:1987} that the  description of such transitions phenomenologically can be done by expanding the free energy up to eighth order in changing physical quantity.  For the case of UGe$_2$  we expand the free energy up to eight order in magnetisation. The properties of such expansion  at zero pressure are explained and applied to UGe$_2$ in \cite{Shopova:2020; Shopova:2021}, but here we include in our considerations also the influence of pressure  and give qualitative results as well as some ideas for further generalisation of the model.\\

 \section{\label{1} Analysis of pressure dependent expansion of  free energy}

\begin{figure}[!ht]
\begin{center}
\includegraphics[scale=0.55]{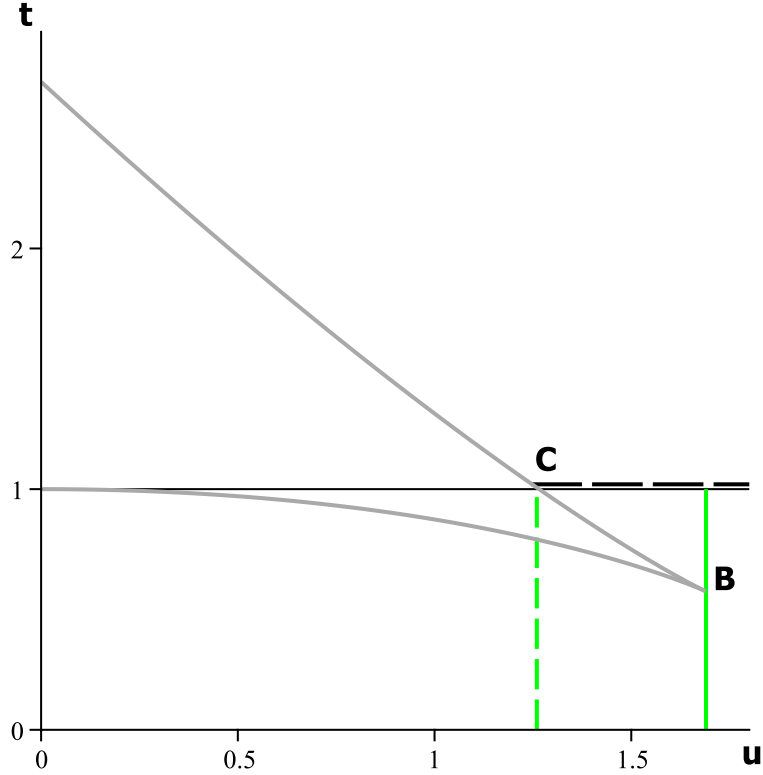}\\
\end{center}
 \caption{\small{Solutions of (\ref{Eq8}) as function of parameter $u$ for $w$=-2.252}}
\label{Fig1}
\end{figure}

  UGe$_2$ has orthorhombic crystal structure of ZrGa$_2$-type with space group \textit{Cmmm} and highly anisotropic magnetisation of Ising type. This is due to localized 5f-electrons, the contribution of itinerant f-electrons to bulk magnetisation is small of order $\approx $ 0.02$ \mu_B$ at ambient pressure as found experimentally. We denote the highest magnetisation  along the easy magnetisation a-axis by  $M_a$; it comes from both itinerant and localised 5f-electrons. The perpendicular components of magnetic moment ($M_b$, $M_c$) of itinerant 5f-electrons along b- and c- crystal axes, should also be considered but due to their smallness compared to $M_a$ , they will be omitted at this stage of calculations.\\
 The thermodynamic analysis is made by using the expansion of free energy in magnetisation up to 8-th order. Such expansion allows the description of so-called isostructural transitions between two ordered phases in which the crystal or magnetic structure are not changed , and only some other physical quantity , for example, magnetisation  as in UGe$_2$ undergoes a change. The properties of the model are explained in detail  in previous papers \cite{Shopova:2020; Shopova:2021} at ambient pressure. Here we will continue the analysis of free energy expansion, taking into account the influence of pressure on the magnetic transitions and thermodynamic quantities.\\
 The free energy density  $F\left[M(T,P)\right]/V=g$ with $V$ the volume, in general form up to 8-th order of  $\overrightarrow{M}=(0,0,M_a)$, is given by the expression; the subscript $a$ will be further omitted.
 \begin{equation}\label{Eq1}
 g=a M^2 +\frac{b}{2}M^4 +\frac{c}{3}M^6+\frac{v}{4}M^8.
\end{equation}
 In the above equation, $a=\alpha \left[T-T_{c}(P)\right]$, where $T_{c}(P)$ is the pressure dependent Curie temperature of second order  transition to FM1, and $\alpha$ is a material parameter. We will also consider that the coefficient $b=b(P)$ may depend on pressure, which dependence will be explained later; for the other two coefficients  $ c, \; v$  at this stage of analysis we assume that they do not depend on temperature and external pressure, namely $;c=c(T_{c0},P=0), \;v=v(T_{c0},P=0)$.  By $T_{c0}=T_{c}(P=0)$ we denote the Curie temperature at  $P=0$ for the transition between paramagnetic phase to FM1. The sign of $b(P), \; c(P=0)$ may be either positive or negative, but $v(P=0)>0$ in order to ensure convergence of $g$.\\
In the expression (\ref{Eq1}) there are too many unknown parameters, which in principle may be derived from experiment and/or respective microscopic calculations in zero external magnetic field. We will not make such quantitative comparison, rather we will give the general qualitative picture of magnetic transitions in UGe$_2$. \\
To reduce the number of unknown parameters, as well as to make free energy density $g$ dimensionless  we redefine  the order parameter $M$ by introducing:
\begin{equation}\label{Eq2}
  m=v^{1/8}M.
\end{equation}
The free energy density - Eq. (\ref{Eq1}) in dimensionless  form reads:
\begin{equation}\label{Eq3}
  f= rm^2+\frac{u}{4}m^4+\frac{w}{3}m^6+\frac{1}{4}m^8;
\end{equation}
and the coefficients in Eq.(\ref{Eq3}) are related to the initial ones in the following way:
\begin{eqnarray*}
       \nonumber u &=& \frac{b}{v^{1/2}}\\
      \nonumber  w &=& \frac{c}{v^{3/4}}
     \end{eqnarray*}
     The reduced temperature is $r =\beta \left[T/T_{c0}-T_{c}(P)/T_{c0}\right]$ with  $\beta=\alpha T_{c0}/v^{1/4}$ .\\
     Here we make some assumptions on the dependence of Curie temperature on pressure. In the general case such dependence may be written as $(P - P_0)^{1/n}$, see, \cite{Shopova:2009} and the arguments given there. The choice of $P_c$ is somewhat arbitrary as for UGe$_2$ there are two special critical pressures: $P_x\sim 1.20$ GPa where the transition from FM1 to FM2 is of first order, and  $P_c \sim 1.5$ GPa, at which ferromagnetic phase - FM1 disappears through first order phase transition. Here we choose $P_0=P_c$ as far as we work at not very high pressures for the description of magnetic phase transitions only and especially the transition from FM1 to FM2 with the increase of pressure. Then $r$ becomes:
    \begin{equation}\label{Eq4}
    r=\gamma(t-1+p)
    \end{equation}
   where $t=T/T_{c0}$ and $p=P/P_c$; the parameter $\gamma$ will be determined in the process of calculations.
The equation of state $(df/dm)$ :
\begin{equation}\label{Eq5}
2m\left[\gamma(t-1+p)+um^2+wm^4+m^6\right]=0
\end{equation}
has an obvious solution for disordered (paramagnetic) phase $m=0$.\\
The stability condition is given by inequality:
\begin{equation}\label{Eq6}
\frac{d^2f}{dm^2}=2\left[\gamma(t-1+p)+3um^2+5wm^4+7m^6\right]\geq 0
\end{equation}
and there are several methods to resolve this condition in analytical form: see for example ~\cite{Izyumov:1984}. This may be also done numerically by direct substitution of solutions of Eq. (\ref{Eq5}) in the above equation and analysing its positiveness. If more than one solution is stable also a comparison between free energies of respective solutions of Eq. (\ref{Eq5}) should be made in order to find out which one in what domain of reduced temperature is an absolute minimum. \\
The equation of state~(\ref {Eq5}) becomes standard 3-rd order algebraic equation with analytical solutions,  if we make the substitution $x=m^2 \geq 0$, ~\cite{Abramowitz:1964}:
\begin{equation}\label{Eq7}
\gamma(t-1+p)+ux+wx^2+x^3=0.
\end{equation}
The solving of Eq.(\ref{Eq7}) begins with the analyses of quantity $Q$ given below, which  determines the number of its real solutions and the boundaries of their existence:
\begin{equation}\label{Eq8}
Q=\frac{1}{4}\left[\gamma(t-1+p)\right]^2 +\frac{2w}{3}\left(\frac{2w^2}{9}-u\right)\gamma(t-1+p)+\frac{u^2}{27}\left(u-\frac{w^2}{4}\right)
\end{equation}
$Q(t)$ is a quadratic equation with respect to the reduced temperature, and we can find those values of reduced temperature $t$ as function of parameters $u,w$ by which we can easily determine the phase boundaries and the order of magnetic transitions as well as the region of  real non-negative solutions for magnetisation as function of parameters $u$ and $w$.  \\
The solutions of (\ref{Eq8})with respect to t, are :
 \begin{equation}\label{Eq9}
 t_{1,2}=\frac{1}{\gamma}\left[-\gamma p + \frac{uw}{3} + \gamma-\frac{2}{27}w^3 \pm \frac{2}{27} \sqrt{(w^2-3u)^3}\right]
 \end{equation}
 It is obvious that $t_{1,2}$ are real for $w^2\geq 3u$ and $t_0$ for $u=w^2/3u$ is a special point, at which
 $$t_1=t_2=\frac{w^3}{27 \gamma} - p +1$$. The calculations show that such point appears only for $w<0$ and $u\geq 0$.
 At this point all three solutions of Eq.(\ref{Eq7}) are equal and real.  As shown before \cite{Shopova:2020; Shopova:2021}, this can be interpreted as the crossover temperature $t_{0cr}$ for the transition between FM1 and FM2 at zero pressure, namely in new notations $t_{0cr}=t_1(p=0)=t_2(p=0)=w^3/27 \gamma +1$ . The expression for reduced temperature $t_{0cr}$ of crossover transition at $p=0$ can be used to find the relation between $\gamma$ and $c$ with the help of respective experimental values of $T_c=$52 K and $T_x=$ 30 K at $P=0$. The numerical result is $\gamma=-26/297w^3$.
We illustrate the dependence of $t_{1,2}$ on parameter $u$ for fixed $w<0$ in Fig. \ref{Fig1}.\\
The grey lines give the variation of $t_{1,2}$ with parameter $u$ at $p=0$. We are mainly interested in the interval of $u$-values between the two vertical green lines, as there a second order phase transition from paramagnetic phase is possible, given by horizontal thick dash line, denoting $t_{cc0}$  Along the thick green line, drawn at $u=w^2/3$ the transition at $t_{c0}$ to FM1 is of second order and at point B a crossover occurs  to FM2 phase with a jump of magnetisation.For $t_2<t<t_1$ , all roots of (\ref{Eq7}) are real and this is the region of first order phase transitions. For $t>t_1$ there is only one real solution and along thick green line above point B, this is the region of phase FM1. Below point B, $t<t_2$ again only one root is real and bigger in magnitude - this is the region of FM2.   Point C ($u=w^2/4$) is tricritical point, at which the transition from disordered phase changes from first to second order.
Further we will show how this picture changes with inclusion of pressure.

\section{\label{3} Results and discussion}
As experiment shows, see for example \cite{Tateiwa:2018} and the papers cited therein, the transition between FM1 and FM2 persists to be of crossover type up to high pressures of order $P_x\sim$ 1.2 GPa, where a first order phase transition occurs  and FM2 disappears. On the other hand the positive muon spin rotation measurements in paper~\cite{Sakarya:2010} give that the transition between FM1 and FM2 becomes a real thermodynamic phase transition at lower pressures  $\sim$ 1 GPa. In recent paper \cite{Liu:2020},  a two-fluid model for magnetic transitions in UGe$_2$ is proposed on the basis of Hall resistivity measurement, where the existence of first order transition in FM2 is questioned.\\
As the conclusions from different experiments are somewhat controversial, here we accept that the transition to FM2 is of  crossover-type  up to some limiting pressure followed by real thermodynamic transition probably of first order through which FM2 disappears at $\sim$ 1.2 GPa.
The free energy expansion we use is a general phenomenological approach to depict transitions between two ordered phases without a change of their structure, and even in its simplified version with respect to influence of pressure on such transitions, it gives indications that the above picture may be realised. \\
For crossover value of $u$, namely, $u=w^2/3$, the equation of state Eq.(\ref{Eq7}) has solutions for $x=M^2$ given by simple formula:
 \begin{eqnarray}
 x_1 &=& \left[\frac{w^3}{27}-\gamma (t-1+p)\right]^{\frac{1}{3}} -\frac{c}{3} \\
 x_{2,3}&=& -\frac{1}{2} \left[\frac{w^3}{27}-\gamma (t-1+p)\right]^{\frac{1}{3}} -\frac{c}{3}\pm i\frac{\sqrt{3}}{2}\left[\frac{w^3}{27}-\gamma (t-1+p)\right]
 \end{eqnarray}
 The calculations give that the smallest root $x_3$ describes the magnetisation of FM1 phase with a second order phase transition at $t_c=1-p$; the jump of magnetisation at crossover temperature $t_{0cr}$ is given by the solution $x_1$ and at the crossover temperature the magnetisation $m^2= x_1=x_2=x_3=-c/3$. We should mention here that the expression for crossover temperature is not changed, when  dependence on pressure is included only in $t_c(p)$, and the other coefficients in the free energy remain independent of $p$.\\In figure below we illustrate this behaviour of $m^2$ for pressure $p=0.1$, which corresponds to $P$ = 0.15 GPa.
 \begin{figure}[!ht]
\begin{center}
\includegraphics[scale=0.55]{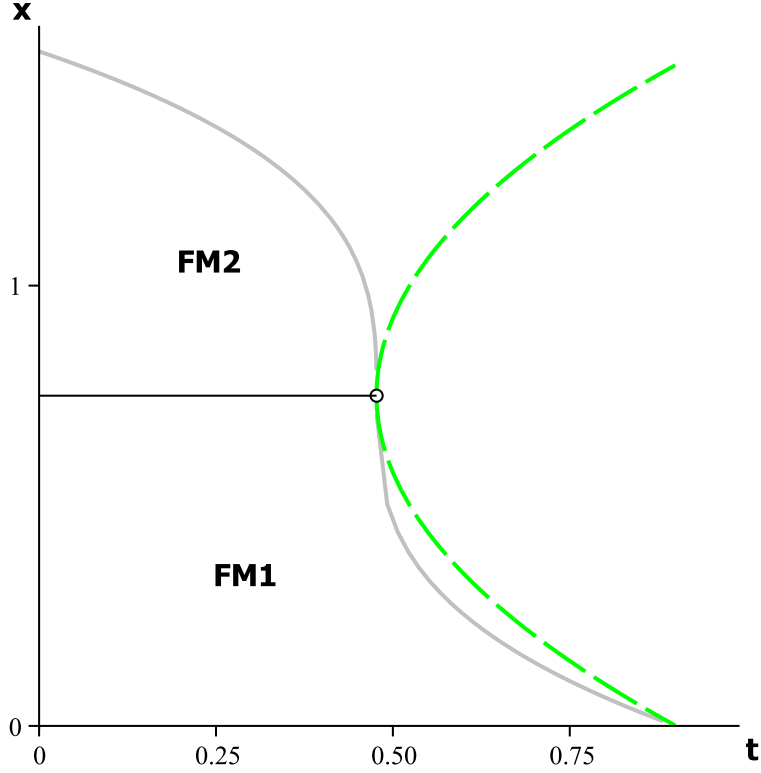}\\
\end{center}
\caption{Function $M^2=x$ of reduced temperature $t$ for pressure $p=0.1$, i.e., $P=$ 0.15 GPa. Open circle denotes the crossover transition between FM1 - lower curve to FM2 - upper curve. The meaning of green lines is explained in the text}
\label{Fig2}
\end{figure}
 The green lines  there show the values of $x=M^2$ which outline the stability of solutions of equation of state. They are found using  the stability condition Eq.(\ref{Eq6}), together with the equation of state Eq.(\ref{Eq5}). The result is  simple quadratic equation in $x$:
 $$\gamma (t-1+p) +\frac{2}{3}w^2 x+wx^2\leq 0.$$
 The roots of this stability condition are depicted in Fig.\ref{Fig2} by green lines, the solutions for $M^2=x$ are stable outside the green lines, note that $w<0$.\\
 In Fig. \ref{Fig3} the behaviour of magnetisation with temperature for different pressures is presented with the aim to show and discuss the limits of approach we apply. It is seen from figure that both temperature of second order transition to FM1, and the temperature of crossover transition decrease with pressure increasing. The region of FM1 existence is also increasing with respect of  FM2 phase. But taking into account that only Curie temperature depends on pressure in linear form immediately leads to some discrepancies, as far as the magnetisation at the crossover temperature remains the same in magnitude as the magnetisation at zero pressure. This is so because we fix the parameter $u$, see Eq.(\ref{Eq5}) to its value $u=w^2/3$ at $p=0$.\\
  \begin{figure}[!ht]
\begin{center}
\includegraphics[scale=0.45]{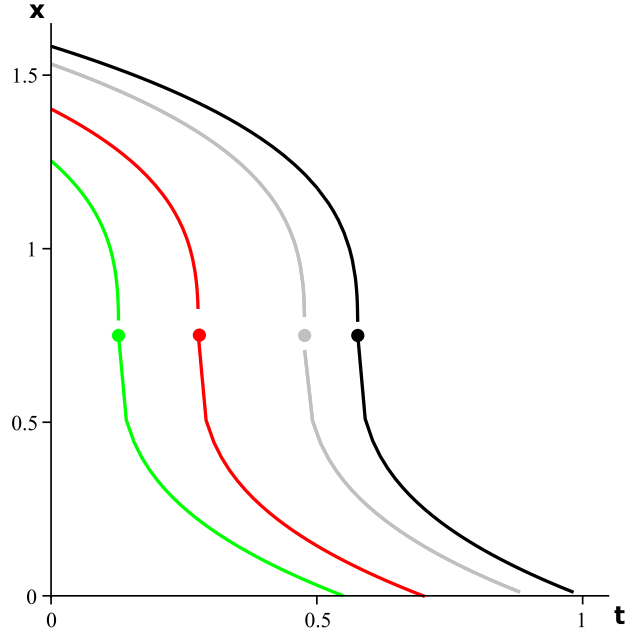}\\
\end{center}
\caption{\small{Function $M^2=x$ of reduced temperature $t$ for different values of pressure: black line is at $p=0$; the grey line is at $p=0.1$, i.e., $P=$ 0.15 GPa; the red line is for $p=0.3$, i.e.,$P=$ 0.45 GPa; and the green line is ar $p=0.45$,i.e., $P=$ 0.675 GPa;   Solid circles of respective color denote the crossover transition.}}
\label{Fig3}
\end{figure}
We suppose that in order to give more realistic picture of FM1 to FM2 transitions and especially to properly describe the behaviour of magnetisation with temperature and pressure, we should assume that $u=u(p)$. The problem with particular form of such dependence is not trivial, as we should take in consideration that the transition from FM1 to FM2 remains a crossover up to pressures $\sim 0.95 - 1$ GPa, and only for bigger pressures a first order phase transition occurs with disappearing of FM2.\\
Even in our simplified assumption there are indications that such transformation is taking place. This is illustrated in Fig. \ref{Fig4}. As we pointed in Section 2, the number of real solutions for magnetisation of equation of state  Eq.(\ref{Eq5}) are determined by the quantity $Q $, Eq.(\ref{Eq8}). Solving it with respect to reduced temperature shows also the limits of existence and gives a picture of possible phase transitions for free energy  Eq.(\ref{Eq3}).\\
   \begin{figure}[!ht]
\begin{center}
\includegraphics[scale=0.45]{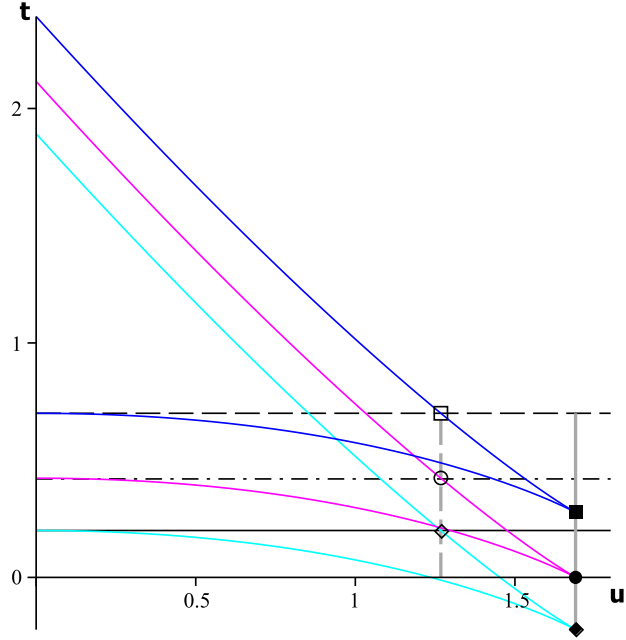}\\
\end{center}
\caption{\small{The dependence of solutions $t_{1,2}$ of Eq.(\ref{Eq8}) on the parameter $u$ for different pressures. Blue lines are drawn for $p=0.3$,i.e.,$P=$ 0.45 GPa. Pink lines are given at limiting pressure $p_l=0.57$, $P=$ 0.86 GPa, and light blue lines are drawn at $p_x=0.8$, i.e.,$P_x=$ 1.2 GPa.  The horizontal black lines denote the reduced Curie temperature $t_c=1-p$ for the respective values of p}}
\label{Fig4}
\end{figure}

In Fig.\ref{Fig4} we have given  solutions  of $Q$ with respect to $t$  as function of parameter $u$ for different pressures. The crossover temperature $t_{cr}$ is determined by the equality $t_1=t_2$, see Eq.(\ref{Eq9}) and the expression below it. It is seen from the figure that $t_{cr}$ is decreasing with pressure increase and above some limiting pressure, which we denote by $p_l$, the point, where $t_1=t_2$ becomes negative and not physical, as $t$ should be $\geq0$. This demonstrates that for pressures $p_l<p<p_x$, our description is too simplified.\\
  We denote the limiting pressure as $p_l$, given by
  $$p_l=1+\frac{w^3}{27 \gamma}.$$
  In Fig 4, $t_{l}^{1,2}$ for $p=p_l$ as function of $u$ are described by  violet lines and the crossover temperature lies on u-axis; it is denoted by solid black circle. The circle on the upper  violet line marks the crossing of $t_1$ with line for second order transition $1-p_l$. The grey vertical lines have the same meaning as the green lines in Fig.\ref{Fig1}. It is obvious that at $p=p_l$ the transition from FM1 to FM2 remains of crossover type.\\
  For pressures $p_l<p<p_x$, the transition cannot be of crossover type any more. This is shown by light blue lines, depicted at $p=p_x$. The crossover temperature within our simplified approach is negative, shown by solid black diamond in Fig.\ref{Fig4}. The values of $u$ for which second order phase transition to FM1 are limited between the crossing of upper blue line and u-axis and the solid grey line. Experimentally, the first order transition between paramagnetic phase and FM1 occurs at pressure $p\sim 0.95$, corresponding to $P \sim$ 1.42 GPa and parameter $u$ should be such as to give such opportunity. Fig. \ref{Fig4} shows that including the dependence of $u$ on pressure is necessary for the proper description of crossover transformation to real first order phase transition. Preliminary calculations show that if $u(P)$ is expanded around the value of $u_0=w^2/3$ at $p=0$ and only linear terms in pressure are preserved, the crossover temperature has proper behaviour, namely, it slightly decreases with increasing of pressure, but another problem appears related to very fast decrease of Curie temperature with pressure and the discrepancy with the respective experimental data, see \cite{Tateiwa:2018}, is of order $6-10$ K. This discrepancy grows with the pressure increase. Moreover, such linear dependence of $u(p)$ on pressure is adequate for the description of transition from paramagnetic phase to FM1, followed by crossover transition to FM2 only for pressures $p<p_l$.\\

\section{\label{4}  Concluding remarks}
In this paper we  apply the phenomenological approach to UGe$_2$ for the description of magnetic phase transitions using the expansion of free energy up to eight order in magnetisation and taking into account in explicit form the dependence on pressure. At this stage we include such dependence only through the linear dependence of Curie temperature  on P. Even in such simplified form, the results show that for pressures smaller that some limiting pressure, defined in the previous section, the transition of FM1 to FM2 remains of crossover type as experimentally found.\\
As pointed above  in order to give more precise picture of magnetic transitions in UGe$_2$ for our model it will be necessary to consider the coefficient in front of fourth order term $u$ in the free energy expansion, Eq. (\ref{Eq3}), as pressure dependent. The concrete form of such pressure dependence should be derived taking into account the available theoretical and experimental data.\\
As it is generally accepted that the spin fluctuations  of itinerant f-electrons are responsible for occurrence of superconductivity of p-type, it will be of great interest also to consider the influence of itinerant f-electrons on magnetic transitions within this phenomenological approach we use. The above issues will be the subject of further studies.

\section*{Acknowledgements}
This work is supported by Grant $KP-06-N38/6$ of the Bulgarian National Science Fund.

\end{document}